\documentclass[twocolumn,showpacs,amsmath,amssymb]{revtex4}

\usepackage{graphicx}
\usepackage{dcolumn}
\usepackage{bm}

\begin{document}

\title{Gamow-Teller properties of the double $\beta$-decay partners 
$^{116}$Cd(Sn) and $^{150}$Nd(Sm)}
\author{D.~Navas-Nicol\'as and P.~Sarriguren}
\affiliation{Instituto de Estructura de la Materia, IEM-CSIC, Serrano
123, E-28006 Madrid, Spain} 
\date{\today}

\begin{abstract}

The two Gamow-Teller (GT) branches connecting the double-beta decay partners
($^{116}$Cd, $^{116}$Sn) and ($^{150}$Nd, $^{150}$Sm) with the intermediate nuclei
$^{116}$In and $^{150}$Pm are studied within a microscopic approach based 
on a deformed proton-neutron quasiparticle random-phase approximation 
built on a Skyrme selfconsistent mean field with pairing correlations and
spin-isospin residual forces. The results are compared with the experimental 
GT strength distributions extracted from charge-exchange reactions.
Combining the two branches, the nuclear matrix elements for the 
two-neutrino double-beta decay are evaluated and compared to 
experimental values derived from the measured half-lives.

\end{abstract}

\pacs{23.40.Hc, 21.60.Jz, 27.60.+j, 27.70.+q}

\maketitle

\section{Introduction\label{sec:introduction}}

Neutrinoless double-beta decay ($0\nu\beta\beta$) is nowadays a topical 
issue that has attracted a lot of interest from both theoretical and 
experimental sides \cite{2breview}. This decay mode is still unobserved 
and would violate lepton number conservation. Its existence would 
demonstrate that the neutrino is a massive Majorana particle and would 
provide a measurement of the absolute mass scale of the neutrino. 
Obviously, the nuclear matrix element (NME) involved in such a process 
must be determined 
accurately to extract a reliable estimate of the neutrino mass. On the 
other hand, the two-neutrino double-beta decay mode ($2\nu\beta\beta$) 
is perfectly allowed by the Standard Model. It is a rare second-order 
weak-interaction process that has been observed experimentally in 
several nuclei (see Ref. \cite{barabash10} for a review). Thus, 
to test the reliability of the nuclear structure calculations involved 
in the $0\nu\beta\beta$ process, one checks first the ability of the 
nuclear models to reproduce the experimental information available.

Certainly, there are differences between the NMEs involved in both 
processes, but there are also clear similarities. Among the differences, 
one should notice that the $2\nu\beta\beta$-decay NMEs are dominated by 
Gamow-Teller (GT) transitions connecting the initial and final $0^+$ states 
with all $J^\pi =1^+$ states in the intermediate nucleus. On the other 
hand, in the $0\nu\beta\beta$ process the intermediate states run over 
all $J^\pi$ values. Another difference with consequences in the 
evaluation of the NMEs is that in $0\nu\beta\beta$ decay the average 
virtual neutrino momentum is much larger than the typical scale of 
nuclear excitations and then closure approximation can be safely used. 
In the case of $2\nu\beta\beta$ decay the neutrino momenta are comparable 
with the nuclear excitation energies, preventing the use of closure.
Among the similarities, the two processes connect the same initial and 
final nuclear ground states to be described within a given nuclear model 
and they share the calculation for the intermediate $J^\pi =1^+$ states.
Therefore, reproducing the $2\nu\beta\beta$ NMEs is a requirement for any 
nuclear structure model aiming to describe the neutrinoless mode.
One can go even further and compare the calculations, not only with the 
$2\nu\beta\beta$ NMEs extracted from the measured half-lives, which is 
nothing but a number, but also with the GT$^-$ (GT$^+$) strength distribution 
of the single branch connecting the initial (final) ground state with all 
the $J^\pi =1^+$ states in the intermediate nucleus. This comparison 
provides a more detailed information on which the calculation can be 
tested. GT strength distributions have been measured in recent years 
from high resolution charge-exchange reactions (CER), within a large 
experimental program aimed to explore the GT properties at low excitation 
energies of double $\beta$-decay 
partners \cite{fujita11,frekers13,rakers05,sasano12,guess11}. 

The purpose of this work is to investigate the possibility to describe the 
rich information available at present on the GT nuclear response within a 
formalism based on a deformed proton-neutron quasiparticle random-phase 
approximation (QRPA). This information includes global properties such as 
the location and strength of the GT resonance, more detailed description in 
the low-lying excitations, and last but not least, the $2\nu\beta\beta$-decay 
NMEs. This study was started for the double $\beta$-decay partners with 
$A=76,128,130$ \cite{sarri12,sarri13}. Here, we shall focus on the 
$^{116}$Cd~$\rightarrow\,^{116}$Sn and $^{150}$Nd~$\rightarrow\,^{150}$Sm 
decays, motivated by the recent experiments on CER
performed for $^{116}$Cd$(p,n)^{116}$In and $^{116}$Sn$(n,p)^{116}$In 
\cite{sasano12}, as well as for $^{150}$Nd$(^3$He,$t)^{150}$Pm and  
$^{150}$Sm$(t,^3$He)$^{150}$Pm \cite{guess11}. It is also worth mentioning 
the recent measurement of the electron capture in $^{116}$In \cite{wrede13} 
that can be used as a benchmark for nuclear structure calculations.
In addition, the double $\beta$ decay of $^{150}$Nd has received increasing
attention in the last years and it is currently considered as one of the
best candidates to search for the $0\nu\beta\beta$ decay in the planned
experiments SNO+, SuperNEMO, and DCBA. The reason for the interest in  
$^{150}$Nd is that it has a large phase-space factor and therefore a 
relatively short half-life. It also has a large $Q_{\beta\beta}$ energy 
that will reduce the background contamination. However, both $^{150}$Nd 
and $^{150}$Sm are deformed nuclei that require a deformed formalism 
to deal with them properly. 

The paper is organized as follows: In Sec. \ref{sec:theory}, we present a
brief summary of the theoretical approach used to describe the
GT strength distributions, as well as the $2\nu\beta\beta$-decay basic
expressions. Section \ref{sec:results} contains the results obtained 
from our approach, which are compared to experimental data and
other available calculations. Section \ref{sec:conclusions} contains a
summary and the main conclusions.

\section{Theoretical approach}
\label{sec:theory}

The description of the deformed QRPA approach used in this work is given 
elsewhere \cite{sarriguren,sarri_pp,alvarez04}. Here we give only a summary 
of the method. We start from a selfconsistent deformed Hartree-Fock (HF) 
calculation with effective nucleon-nucleon Skyrme interactions, assuming 
axial deformation and time reversal symmetry \cite{vautherin}. 
Most of the results in this work are performed with the Skyrme 
force SLy4 \cite{chabanat98}, which is one of the most widely used
and successful interactions. In addition, we also present results obtained 
with other Skyrme forces to investigate their suitability for
the description of the spin-isospin properties of nuclei. In particular,
besides SLy4, we shall show results from the simpler force Sk3 \cite{sk3} 
and from the force SGII \cite{sg2} that was fitted taking into account 
nuclear spin-isospin properties.

\begin{figure}[htb]
\begin{center}
\includegraphics[width=8.5cm]{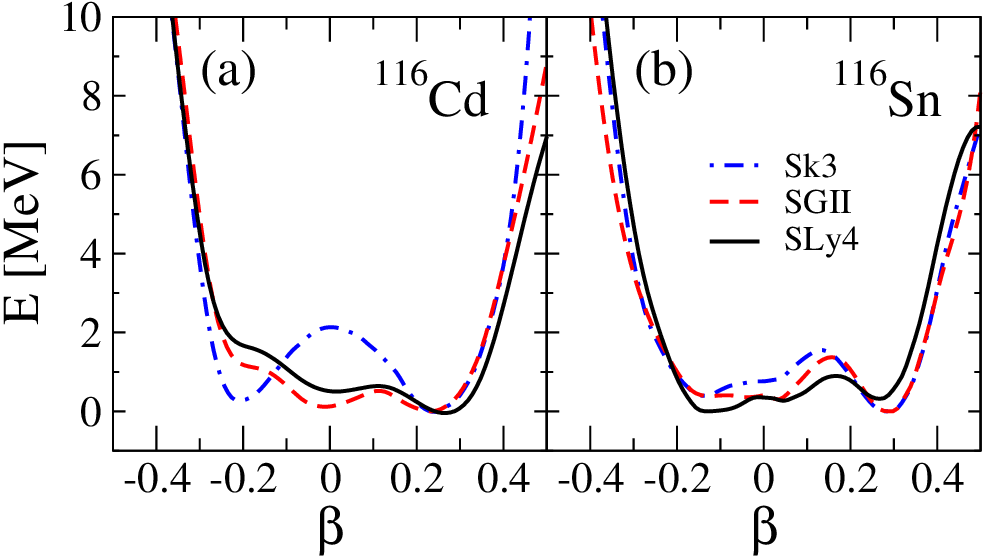}
\caption{(Color online) 
Energy-deformation curves for $^{116}$Cd (a) and $^{116}$Sn (b) obtained 
from HF+BCS calculations with various Skyrme forces.}
\label{fig1}
\end{center}
\end{figure}

\begin{figure}[htb]
\begin{center}
\includegraphics[width=8.5cm]{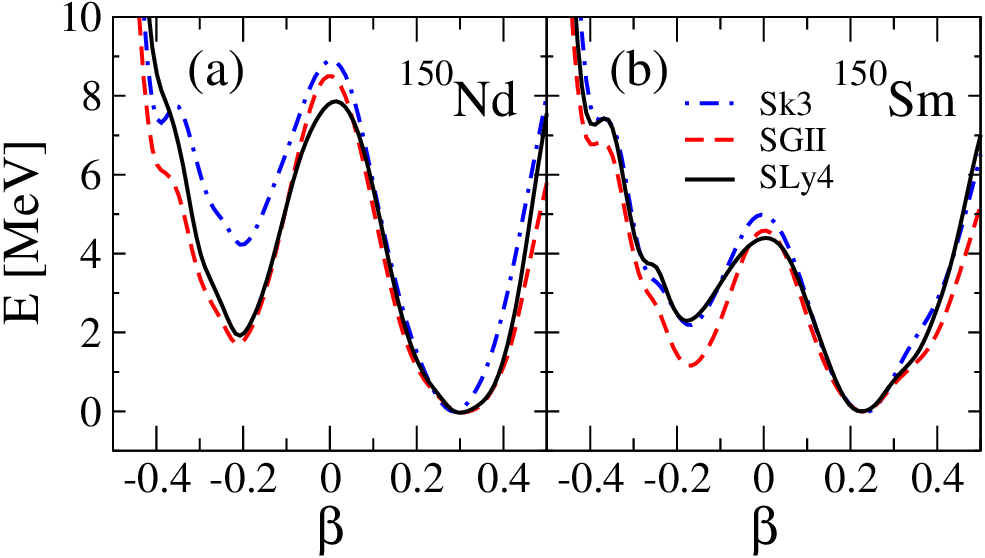}
\caption{(Color online) Same as in Fig.~\ref{fig1}, but for $^{150}$Nd (a) 
and $^{150}$Sm (b).}
\label{fig2}
\end{center}
\end{figure}

In our approach, the single-particle wave functions 
are expanded in terms of the eigenstates of an axially symmetric harmonic 
oscillator in cylindrical coordinates using twelve major shells. Pairing 
correlations between like nucleons are included in BCS approximation
taking fixed pairing gap parameters for protons and neutrons, which are
phenomenologically determined. Besides the selfconsistent HF solution, 
we also explore the potential energy curves, that is, the HF energy as 
a function of the quadrupole deformation $\beta$, which are obtained 
from constrained HF+BCS calculations. These results can be seen in 
Fig. \ref{fig1} for $^{116}$Cd and $^{116}$Sn, and in Fig. \ref{fig2} 
for $^{150}$Nd and $^{150}$Sm for the three Skyrme forces considered.
We get very soft profiles for $^{116}$Cd and $^{116}$Sn, whereas for 
$^{150}$Nd and $^{150}$Sm we obtain two energy minima, oblate and prolate,
but with clear prolate ground states in both cases. We obtain similar 
results with the three Skyrme forces with the only exception of the 
force Sk3 in $^{116}$Cd, which seems to miss a spherical solution.

\begin{figure}[htb]
\begin{center}
\includegraphics[width=8.5cm]{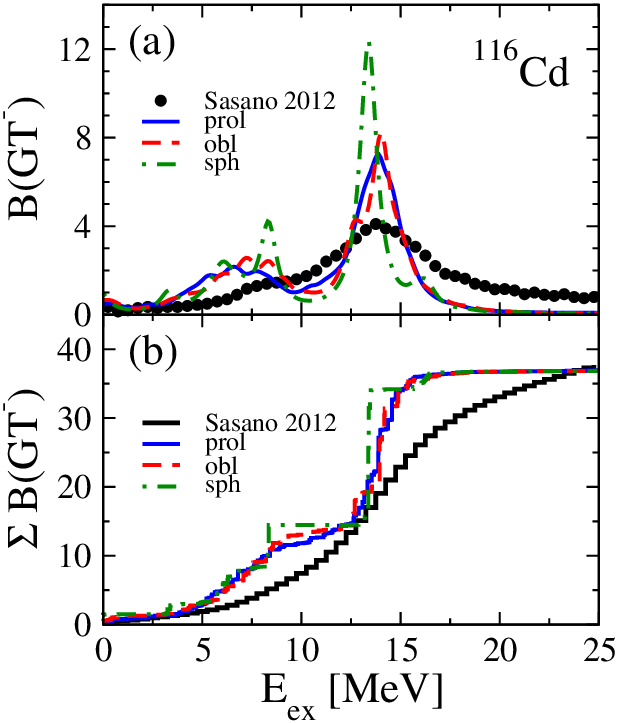}
\caption{(Color online) GT strength distributions $B({\rm GT}^-)$  
in $^{116}$Cd as a function of the excitation energy in the daughter nucleus 
(a) and accumulated strengths (b). 
SLy4-QRPA calculations for various deformations are compared to data 
(Sasano 2012: \cite{sasano12}) from $(p,n)$ reactions.}
\label{fig3}
\end{center}
\end{figure}

\begin{figure}[htb]
\begin{center}
\includegraphics[width=8.5cm]{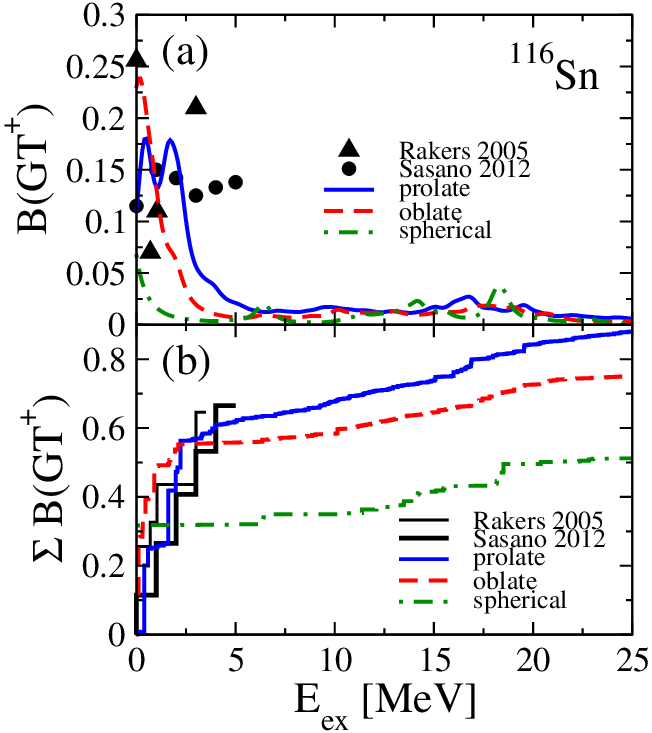}
\caption{(Color online) GT strength distributions $B({\rm GT}^+)$ 
in $^{116}$Sn  as a function of the excitation energy in the daughter nucleus 
(a) and accumulated strengths (b).
SLy4-QRPA calculations for various deformations are compared to data 
(Sasano 2012: \cite{sasano12}) from $(n,p)$ reactions, as well as from 
$(d,^2$He) (Rakers 2005: \cite{rakers05}).}
\label{fig4}
\end{center}
\end{figure}

\begin{figure}[htb]
\begin{center}
\includegraphics[width=8.5cm]{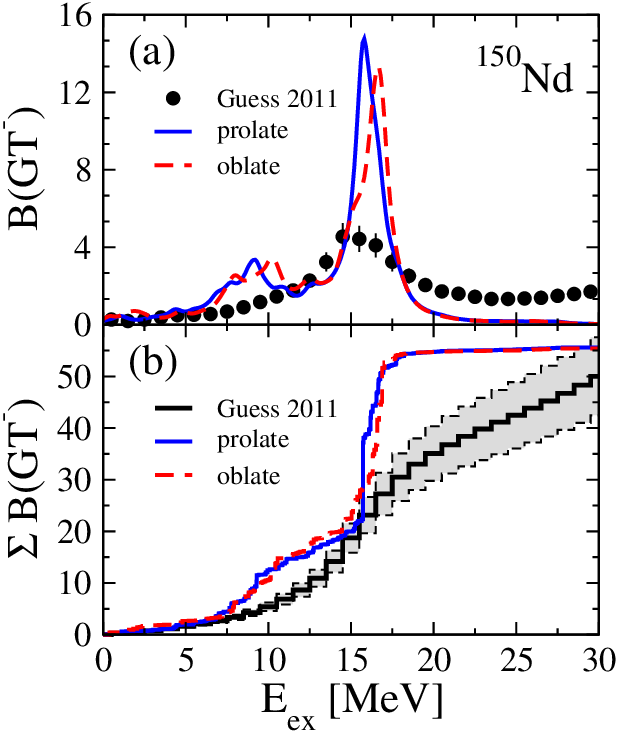}
\caption{(Color online) Same as in Fig. \ref{fig3}, but for $B({\rm GT}^-)$ 
in $^{150}$Nd. Data (Guess 2011: \cite{guess11}) are from $(^3$He,t) CERs.}
\label{fig5}
\end{center}
\end{figure}

\begin{figure}[htb]
\begin{center}
\includegraphics[width=8.5cm]{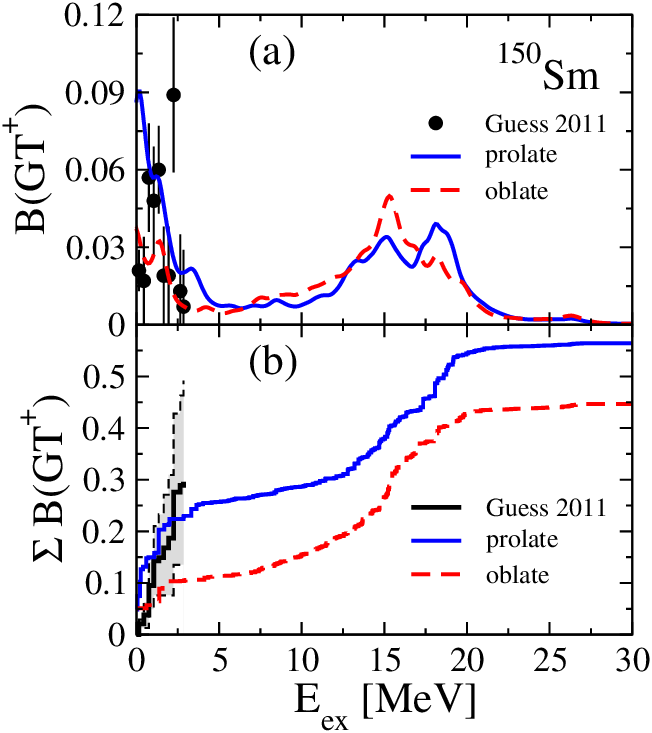}
\caption{(Color online) Same as in Fig. \ref{fig4}, but for $B({\rm GT}^+)$ in 
$^{150}$Sm. Data (Guess 2011: \cite{guess11}) are  from (t,$^3$He) CERs.} 
\label{fig6}
\end{center}
\end{figure}

\begin{figure}[htb]
\begin{center}
\includegraphics[width=8.5cm]{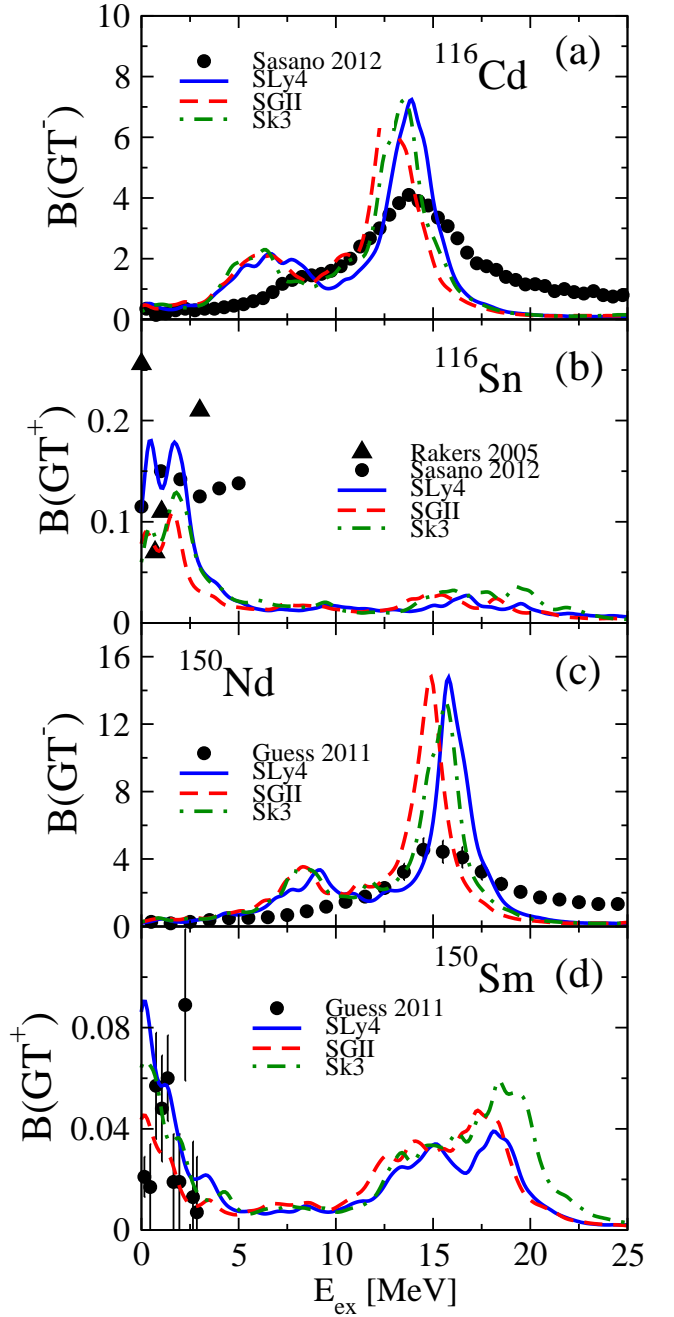}
\caption{(Color online) GT strength distributions in $^{116}$Cd (a),
$^{116}$Sn (b), $^{150}$Nd (c), and $^{150}$Sm (d). The results correspond
to the prolate shapes with the Skyrme forces SLy4, SGII, and Sk3.
Data are as in the previous figures.}
\label{fig7}
\end{center}
\end{figure}

\begin{figure}[htb]
\begin{center}
\includegraphics[width=8.5cm]{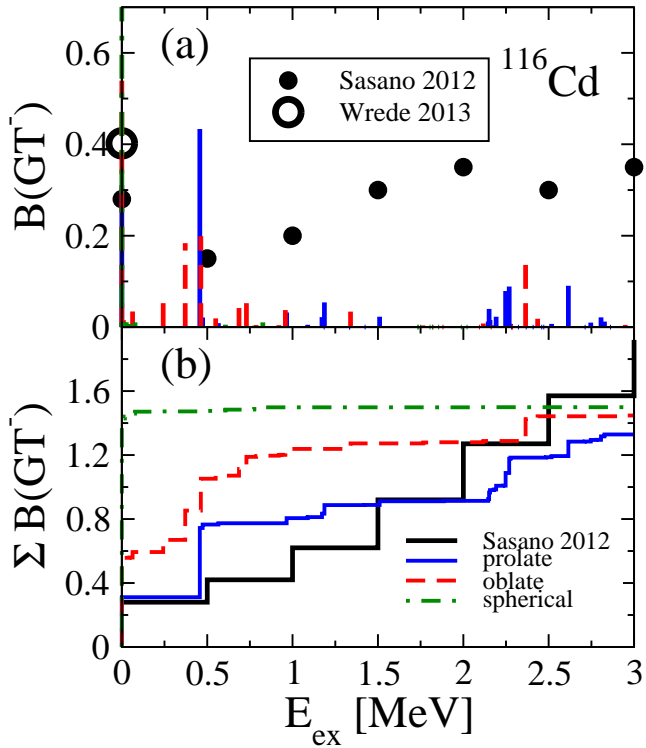}
\caption{(Color online)  Same as in Fig. \ref{fig3}, but in the low
energy range. Also shown is the GT strength extracted from electron
capture of $^{116}$In (Wrede 2013: \cite{wrede13}).}
\label{fig8}
\end{center}
\end{figure}

\begin{figure}[htb]
\begin{center}
\includegraphics[width=8.5cm]{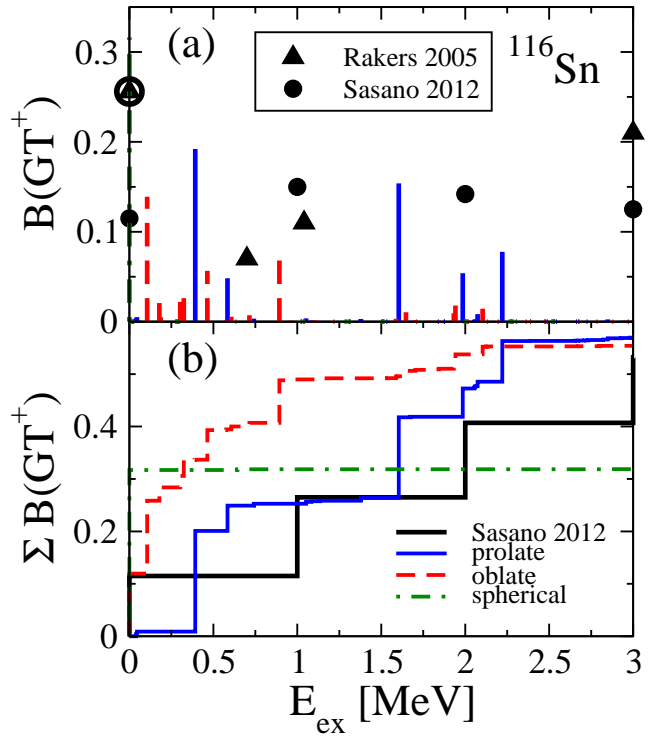}
\caption{(Color online)  Same as in Fig. \ref{fig4}, but in the low
energy range.  Also shown (open circle) is the GT strength extracted from $\beta^-$
decay of $^{116}$In to which the data from \cite{rakers05} are normalized.}
\label{fig9}
\end{center}
\end{figure}

\begin{figure}[htb]
\begin{center}
\includegraphics[width=8.5cm]{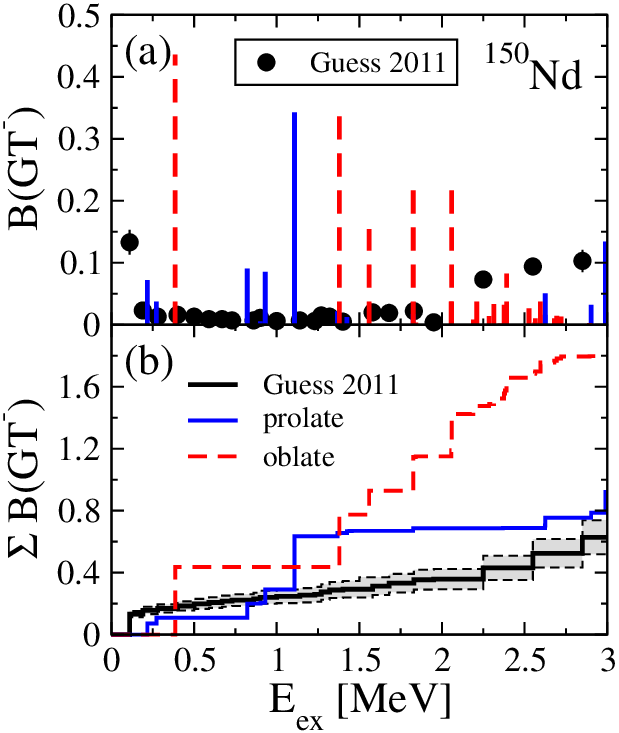}
\caption{(Color online)  Same as in Fig. \ref{fig5}, but in the low
energy range.}
\label{fig10}
\end{center}
\end{figure}

\begin{figure}[htb]
\begin{center}
\includegraphics[width=8.5cm]{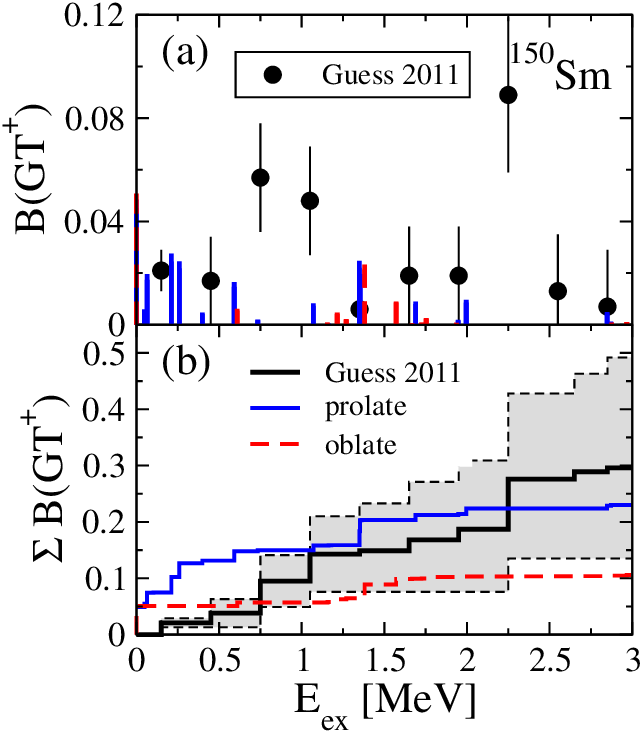}
\caption{(Color online)  Same as in Fig. \ref{fig6}, but in the low
energy range.}
\label{fig11}
\end{center}
\end{figure}

\begin{figure}[htb]
\begin{center}
\includegraphics[width=8.5cm]{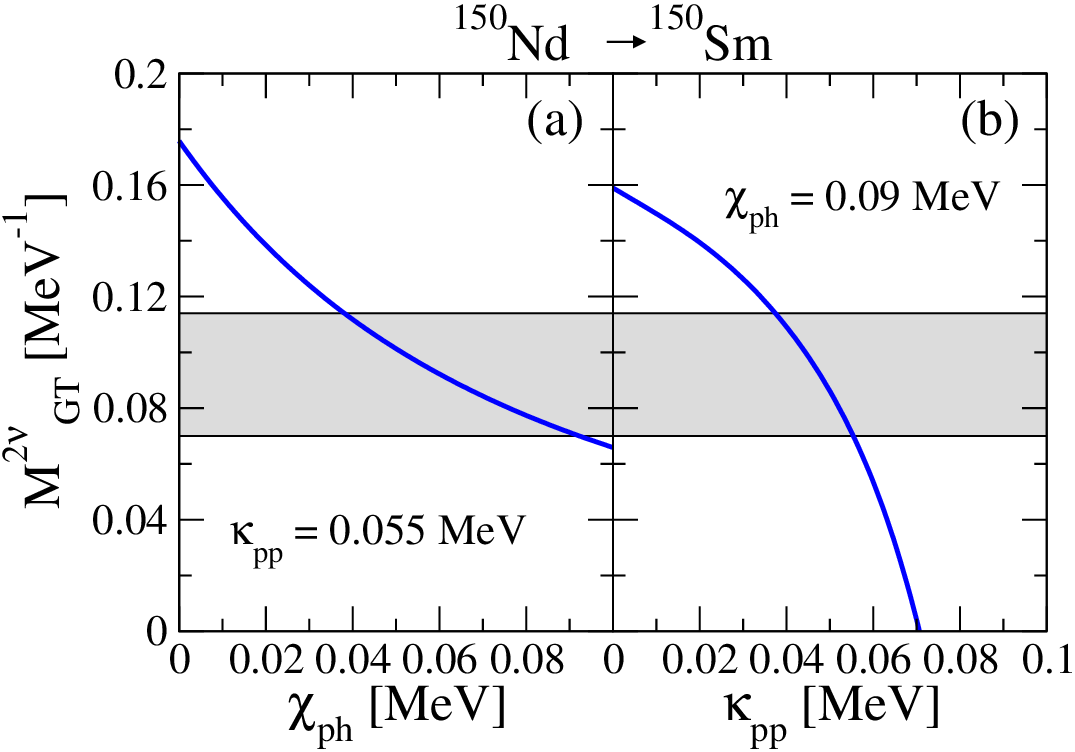}
\caption{(Color online) Nuclear matrix element for the $2\nu\beta\beta$ decay 
of $^{150}$Nd as a function of the coupling strengths $\chi_{ph}^{GT}$ (a) and
$\kappa_{pp}^{GT}$ (b) for prolate shapes in both  $^{150}$Nd and $^{150}$Sm.
The shaded area indicates the experimental range extracted from the 
measured half-life using bare $g_A=1.273$ (lower line) and quenched $g_A=1$ 
(upper line) factors.}
\label{fig12}
\end{center}
\end{figure}

\begin{figure}[htb]
\begin{center}
\includegraphics[width=8.cm]{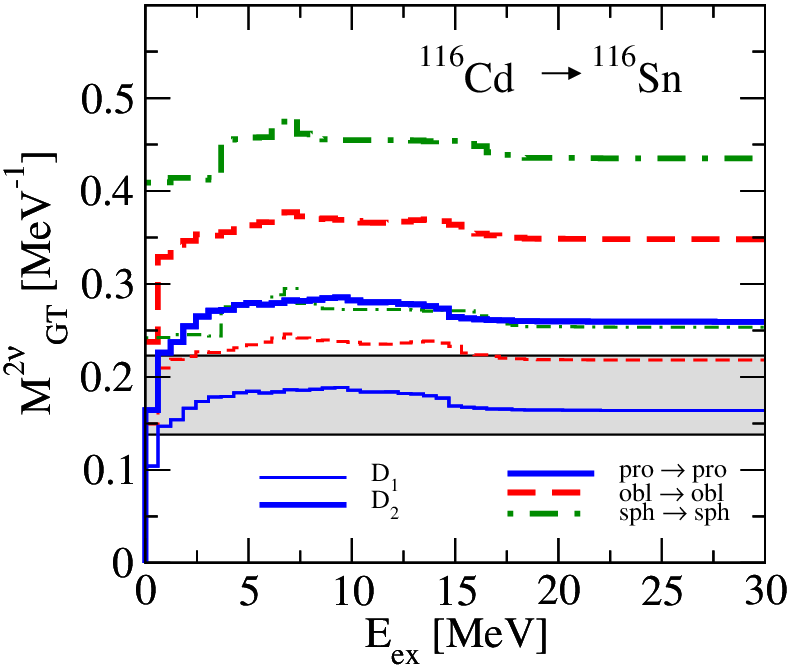}
\caption{(Color online)  Running sums of the $2\nu\beta\beta$ NME in 
$^{116}$Cd as a function of the intermediate 
excitation energy in $^{116}$In for various calculations using different 
quadrupole deformations for parent and daughter nuclei.
Thin lines correspond to calculations with the energy denominator
$D_1$ (\ref{den1}) while thick lines correspond to $D_2$  (\ref{den2}).
}
\label{fig13}
\end{center}
\end{figure}

\begin{figure}[htb]
\begin{center}
\includegraphics[width=8.cm]{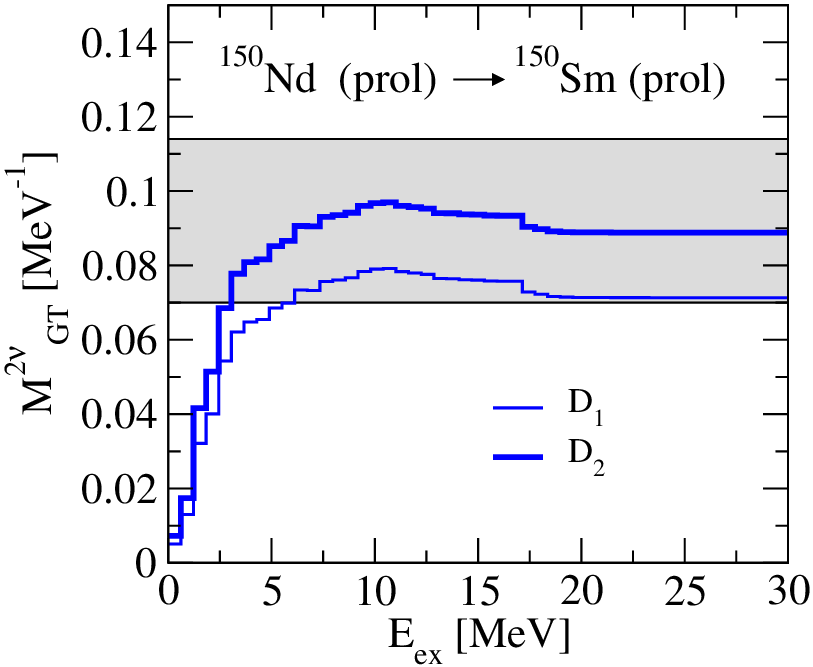}
\caption{(Color online) Same as in Fig. \ref{fig13}, but for the 
double-$\beta$ decay of $^{150}$Nd.}
\label{fig14}
\end{center}
\end{figure}

\begin{figure}[htb]
\begin{center}
\includegraphics[width=8.cm]{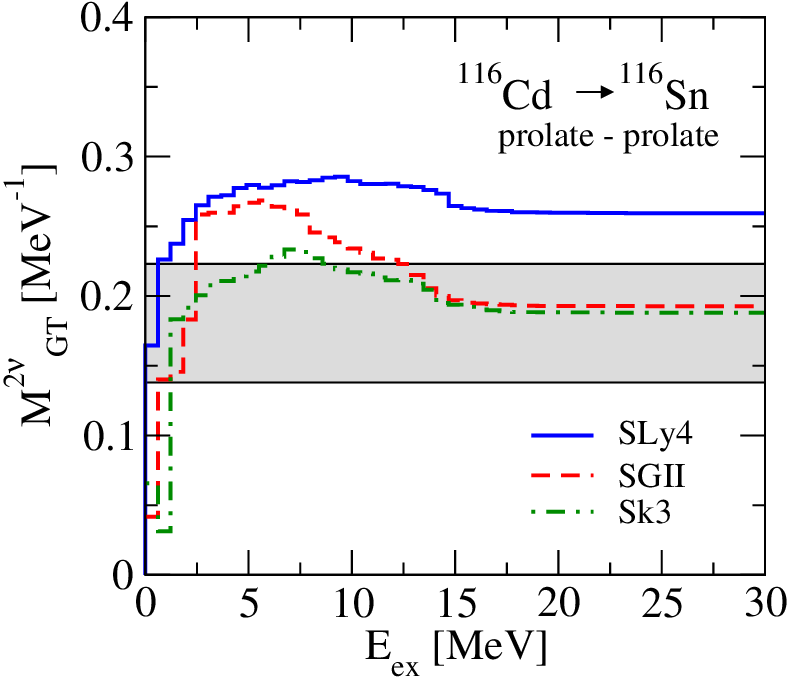}
\caption{(Color online) Running sums of the $2\nu\beta\beta$ NME in 
$^{116}$Cd as a function of the intermediate 
excitation energy in $^{116}$In for the decay between prolate shapes.
The results are obtained with the energy denominator $D_2$ for three 
different Skyrme forces.}
\label{fig15}
\end{center}
\end{figure}

\begin{figure}[htb]
\begin{center}
\includegraphics[width=8.cm]{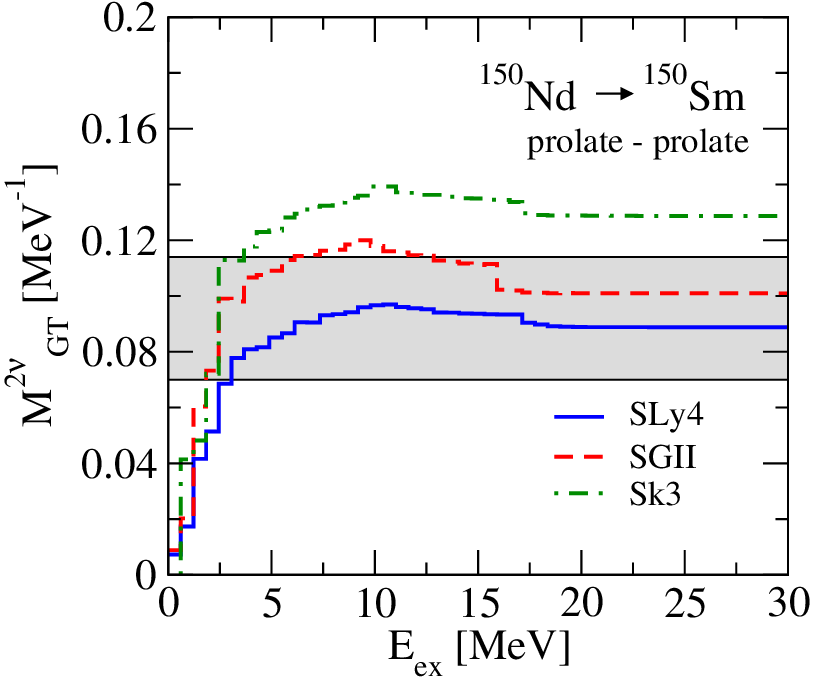}
\caption{(Color online) Same as in Fig. \ref{fig15}, but for the decay
of $^{150}$Nd.}
\label{fig16}
\end{center}
\end{figure}

After the HF+BCS calculation, the QRPA equations are solved on the 
deformed ground-state basis to get the GT strength distributions and to 
compute the $2\nu\beta\beta$-decay NME. To describe GT 
excitations in QRPA we add to the quasiparticle mean field a separable 
spin-isospin residual interaction in the particle-hole ($ph$) and 
particle-particle ($pp$) channels. The $ph$ part is responsible for 
the position and structure of the GT resonance and its coupling constant
$\chi_{ph}^{GT}$ is usually taken to reproduce them
\cite{sarriguren,moller,hir,homma}.  The $pp$ part consists of a 
proton-neutron pairing force and it is also introduced as a separable 
force \cite{sarri_pp,hir}. The coupling constant $\kappa_{pp}^{GT}$ is usually 
fitted to the half-lives phenomenology \cite{homma}. In this work we 
have chosen for the $ph$ and $pp$ coupling constants the law dependence 
on the mass number $A$ given by $C/A^{0.7}$ in Ref. \cite{homma}. 
However, the different mean fields used, Nilsson \cite{homma} versus 
Skyrme-HF here, make the constants somewhat different. Specifically 
we use $\chi_{ph}^{GT}=3.0/A^{0.7}{\rm MeV}$ and $\kappa_{pp}^{GT}=1.8/A^{0.7}$ MeV.
Although slightly different values could be used for the different
Skyrme forces, we use in this work the same values for them, so that 
changes among calculations from various forces are caused by the underlying
quasiparticle structure provided by the effective Skyrme force.
At this point it is worth mentioning the studies on $^{150}$Nd and 
$^{150}$Sm carried out in Refs. \cite{yousef09,fang10}, where deformed 
QRPA calculations were performed using realistic nucleon-nucleon residual 
interactions based on the Brueckner $G$ matrix for the CD-Bonn force 
on top of a phenomenological deformed Woods-Saxon potential. 
Comparison with the results 
obtained from schematic separable forces \cite{yousef09} shows that the 
latter reproduce fairly well the main characteristic of the GT strength 
distributions profiles and $2\nu\beta\beta$-decay NMEs.

The technical details to solve the QRPA equations have been described in
Refs.~\cite{hir,sarriguren,sarri_pp}. For each value of the excitation 
energy, the GT transition amplitudes in the intrinsic frame connecting 
the ground state $\left| 0\right\rangle $ to one-phonon states in the 
daughter nucleus $\left| \omega _K \right\rangle $, are found to be

\begin{equation} 
\left\langle \omega _K | \sigma _Kt^{\pm} | 0 \right\rangle = \mp 
M^{\omega _K}_\pm \, ,
\label{amplgt}
\end{equation}
where 
\begin{eqnarray}
M_{-}^{\omega _{K}}&=&\sum_{\pi\nu}\left( 
v_{\nu}u_{\pi} X_{\pi\nu}^{\omega _{K}}+
u_{\nu}v_{\pi} Y_{\pi\nu}^{\omega _{K}}\right) 
\left\langle \nu\left| \sigma _{K}\right| 
\pi\right\rangle , \\ 
M_{+}^{\omega _{K}}&=&\sum_{\pi\nu}\left( 
u_{\nu}v_{\pi} X_{\pi\nu}^{\omega _{K}}+ 
v_{\nu}u_{\pi} Y_{\pi\nu}^{\omega _{K}}\right) 
\left\langle \nu\left| \sigma _{K}\right| 
\pi\right\rangle \, ,
\end{eqnarray}
in terms of the occupation amplitudes for neutrons and protons $v_{\nu,\pi}$ 
($u^2_{\nu,\pi}=1-v^2_{\nu,\pi}$) and the forward and backward amplitudes of 
the QRPA phonon operator, $X_{\pi\nu}^{\omega _{K}}$ and $Y_{\pi\nu}^{\omega _{K}}$, 
respectively. Once the intrinsic amplitudes in Eq.~(\ref{amplgt}) are 
calculated, the GT strength $B$(GT) in the laboratory frame for 
a transition $I_i K_i (0^+0)\rightarrow I_fK_f(1^+K)$ can be obtained as

\begin{eqnarray}
B_{\omega}({\rm GT}^\pm ) &=& \sum_{\omega_{K}} \left[ \left\langle \omega_{K=0} 
\left| \sigma_0t^\pm \right| 0 \right\rangle ^2 \delta (\omega_{K=0}-
\omega ) \right. \nonumber \\
&& + 2\left. \left\langle \omega_{K=1} \left| \sigma_1t^\pm \right| 
0 \right\rangle ^2 \delta (\omega_{K=1}-\omega ) \right] \, .
\label{bgt}
\end{eqnarray}
To obtain this expression we have used the initial and final states in 
the laboratory frame expressed in terms of the intrinsic states using 
the Bohr and Mottelson factorization. Finally, a quenching factor 
$q=g_{A}/ g_{A,{\rm bare}} = 0.79$ is included in the calculations to 
take into account in an effective way all the correlations
that are not properly considered in the present approach.
We shall discuss later in Subsec. \ref{dbd} various attempts aiming
to fit simultaneously single-$\beta$ and double-$\beta$ decay 
observables by adjusting the quenching factor.

Concerning  the $2\nu\beta\beta$-decay NMEs, the basic expressions for 
this process, within the deformed QRPA formalism used in this work, can 
be found in Refs. \cite{alvarez04,simkovic04,moreno09}. 
Deformation effects on the $2\nu\beta\beta$ NMEs have been also studied
within the Projected Hartree-Fock-Bogoliubov model \cite{singh07}. 
Attempts to describe deformation effects on the $0\nu\beta\beta$ decay
within QRPA models can also be found in Refs. \cite{fang10r,mustonen13}.
The half-life of the $2\nu\beta\beta$ decay can be written as 

\begin{equation}\label{half-life}
\left[ T_{1/2}^{2\nu\beta\beta}\left( 0^+_{\rm gs} \to 0^+_{\rm gs}  
\right) \right] ^{-1}= (g_A)^4\ G^{2\nu\beta\beta}\ \left| 
M_{GT}^{2\nu\beta\beta}\right| ^2 \, ,
\end{equation}
in terms of the phase-space integral $G^{2\nu\beta\beta}$ and the nuclear 
matrix element $M_{GT}^{2\nu\beta\beta}$ that contains all the 
information of the nuclear structure involved in the process,

\begin{eqnarray}\label{mgt}
&& M_{GT}^{2\nu\beta\beta}=\sum_{K=0,\pm 1}\sum_{m_i,m_f} (-1)^K 
\frac{\langle \omega_{K,m_f}  | 
\omega_{K,m_i}  \rangle } {D} \nonumber \\
&& \times \langle 0_f| \sigma_{-K}t^-| \omega_{K,m_f}  \rangle \: 
\langle \omega_{K,m_i}  | \sigma_Kt^- | 0_i \rangle \, .
\end{eqnarray}
In this equation $|\omega_{K,m_i}\rangle (|\omega_{K,m_f}\rangle)$ are the 
QRPA intermediate $1^+$ states reached from the initial (final) nucleus.
The indices $m_i$, $m_f$ label the $1^+$ states
of the intermediate nucleus. The GT matrix elements are those in 
Eq. (\ref{amplgt}). The overlaps are needed to take into account
the non-orthogonality of the intermediate states reached from different 
initial $| 0_i \rangle$ and final $ | 0_f \rangle$ ground states. 
Their expressions can be found in Ref.~\cite{simkovic04}.
The energy denominator $D$ involves the energy of the emitted leptons,
which is given on average by $\frac{1}{2} Q_{\beta \beta}+m_e$, as well as
the excitation energies of the intermediate nucleus.
In terms of the QRPA excitation energies the denominator
can be written as 
\begin{equation}\label{den1}
D_1= \frac{1}{2} (\omega_K^{m_f}+ \omega_K^{m_i}),
\end{equation}
where  $\omega_K^{m_i} (\omega_K^{m_f})$ is the QRPA excitation energy
relative to the initial (final) nucleus.
It turns out that the NMEs are quite sensitive to the values of the
denominator, especially for low-lying states when the denominator
reaches smaller values.
Thus, it is a common practice to use some experimental normalization
of this denominator trying to improve the accuracy of the NMEs.
In this work we shall also consider the denominator $D_2$, which
is corrected with the experimental energy $\bar{\omega}_{1_1^+}$
of the first $1^+$ state in the intermediate nucleus relative to 
the mean ground-state energy
of the initial and final nuclei, in such a way that the first calculated
$1^+$ state appears at the exact experimental energy,

\begin{equation}\label{den2}
D_2= \frac{1}{2} \left[ \omega_K^{m_f}+ \omega_K^{m_i}-\left( \omega_K^{1_f}+ 
\omega_K^{1_i}\right) \right]  
+\bar{\omega}_{1_1^+} \, .
\end{equation}
The running $2\nu\beta\beta$ sums calculated later, will be shown for the
two choices of the denominator $D_1$ and $D_2$. 
In the case of the $A=116$ isobars, the ground state in the intermediate
nucleus $^{116}$In is a $1^+$ state. Then, the  energy 
$\bar{\omega}_{1_1^+}$ is given by 

\begin{equation}\label{wexp}
\bar{\omega}_{1_1^+}=\frac{1}{2}(Q_{EC}+Q_{\beta^-})_{\rm exp} \, ,
\end{equation}
written in terms of the experimental energies $Q_{EC}=0.463$ MeV and 
$Q_{\beta^-}=3.276$ MeV of the decays of $^{116}$In into $^{116}$Cd and 
$^{116}$Sn, respectively.
In the case of $A=150$,  although the ground state in $^{150}$Pm 
is not a $1^+$ state, the first $1^+$ state excited in CERs 
appears at a very low excitation energy, E=0.11 MeV \cite{guess11}.
Hence, we also determine $\bar{\omega}_{1_1^+}$  from 
the experimental values of $Q_{EC}=0.083$ MeV and 
$Q_{\beta^-}=3.456$ MeV for the decays of $^{150}$Pm into $^{150}$Nd and 
$^{150}$Sm, respectively.

The various measurements reported for the  $2\nu\beta\beta$-decay 
half-lives ($T_{1/2}^{2\nu\beta\beta}$)
have been analyzed in Ref.~\cite{barabash10}, where recommended values of 
$ T_{1/2}^{2\nu\beta\beta}(^{116}$Cd)=$(2.8 \pm 0.2)\times 10^{19}$ yr and
$ T_{1/2}^{2\nu\beta\beta}(^{150}$Nd)=$(8.2 \pm 0.9)\times 10^{18}$ yr 
were adopted. Using the phase-space factors from the recent evaluation  
\cite{kotila12} that involves exact Dirac wave functions with finite 
nuclear size and electron screening,
$G^{2\nu\beta\beta}\ (^{116}$Cd)=$2.764 10^{-18}$ yr$^{-1}$, 
and $G^{2\nu\beta\beta}\ (^{150}$Nd)$=3.643 10^{-17}$ yr$^{-1}$, 
one gets the experimental nuclear matrix elements 
$M_{GT}^{2\nu\beta\beta}\ (^{116}$Cd)=0.138 MeV$^{-1}$ (0.223 MeV$^{-1}$)
when the bare  $g_{A,{\rm bare}}=1.273$ (quenched factor $g_A=1$) is used, and
$M_{GT}^{2\nu\beta\beta}\ (^{150}$Nd)=0.070 MeV$^{-1}$ (0.114 MeV$^{-1}$)
when  $g_{A,{\rm bare}}=1.273$ ($g_A=1$) is used.

\section{Results}
\label{sec:results}

\subsection{Gamow-Teller strength distributions}

Next we discuss the results obtained for the energy distributions of the
GT strength.
In the upper panels (a) of Figs. \ref{fig3} and \ref{fig4} we can see the GT 
strength distributions for $^{116}$Cd ($B({\rm GT}^-)$) and 
$^{116}$Sn ($B({\rm GT}^+)$), respectively.
The lower panels (b) show the respective accumulated GT strengths. 
We compare our QRPA results from SLy4 obtained for prolate, oblate, and
spherical shapes in the initial ($^{116}$Cd) and final ($^{116}$Sn) nuclei
with the experimental strengths extracted
from $^{116}$Sn$(d,^2$He)$^{116}$In CERs \cite{rakers05} and from 
$^{116}$Cd$(p,n)^{116}$In and $^{116}$Sn$(n,p)^{116}$In CERs \cite{sasano12}.
One should notice that the measured strength extracted from the cross 
sections contains two types of contributions, namely GT ($\sigma t^\pm$ 
operator) and isovector spin monopole (IVSM) ($r^2 \sigma t^\pm$ operator). 
Both are associated to $\Delta L=0$ and $\Delta S=1$ transitions and 
therefore, they have similar angular distributions that cannot be 
disentangled in the experiment. Thus, the measured strength corresponds 
actually to B(GT+IVSM).
Different theoretical calculations evaluating the contributions from
both GT and IVSM modes are available in the literature
\cite{sasano12,hamamoto00,bes12,civitarese14}.
In Ref. \cite{hamamoto00} a self-consistent Skyrme Hartree-Fock plus 
Tamm-Dancoff approximation was used to separate the GT from 
the IVSM strength. Well separated centroids of both resonances were found, 
with the IVSM resonance at much higher energy, as it corresponds to a 
$2\hbar\omega$ mode. Subsequent calculations including $A=116$ isobars were 
performed in Refs. \cite{bes12,civitarese14} with different
degrees of complexity, from a schematic single-particle model space up 
to realistic calculations within a QRPA approach in a Woods-Saxon
basis and two-body interactions constructed with the Bonn-A potential.
In Ref. \cite{sasano12} the contribution from the IVSM component was also 
evaluated by employing a microscopic method based on QRPA with residual 
interactions obtained from the Brueckner G-matrix for CD-Bonn 
nucleon-nucleon force.
In general, the conclusions of those theoretical calculations converge
in an overall picture. In the $(p,n)$ direction the strength distribution 
is dominated by the GT component up to about 20 MeV, although non-negligible
contributions from IVSM components are found between 10 and 20 MeV, 
overlapping with the GT strength of the GT giant resonance. 
Above this energy, there is no significant amount of GT strength in the 
calculations. The interference effects are constructive below 20 MeV and
destructive above this energy, but they are not very important.
In the $(n,p)$ direction the GT strength is expected to be strongly Pauli 
blocked in nuclei with more neutrons than protons and therefore, the 
measured strength is mostly due to the IVSM resonance over the whole 
excitation-energy range. 
Nevertheless, the strength found in low-lying isolated peaks is associated 
with GT transitions because the continuous tail of the IVSM resonance is 
very small at these energies and is not expected to exhibit any peak.
In summary, the measured strength in the ($p,n$) direction can be safely 
assigned to be GT in the low energy range below 10 MeV and with some 
caution between 10 and 20 MeV. Beyond 20 MeV the strength would be
practically due to IVSM. On the other hand the measured strength in 
the ($n,p$) direction would be due to IVSM transitions, except in the
low-lying excitation energy below several MeV, where the isolated peaks
observed can be attributed to GT strength. This is the reason why we
plot experimental data in Fig. \ref{fig4} only up to 5 MeV.

Calculations for the GT strength distributions in both directions 
B(GT$^{\pm}$) for the $A=116$ isobars are also available from spherical 
QRPA calculations within a Woods-Saxon basis and two-body Bonn-A realistic 
interactions \cite{suhonen14}. The results from this reference are quite 
similar to those in Figs. \ref{fig3} and \ref{fig4}, although a direct 
comparison is not easy as they are obtained for particular values of 
both the strength of the $pp$ residual interaction and the effective 
$g_A$ value. The B(GT$^-$) strength 
distribution calculated in Ref. \cite{suhonen14} is characterized 
by some marginal strength up to an excitation energy of 5 MeV, then
a small bump develops between 7 and 10 MeV, and finally the GT resonance
appears between 11 and 15 MeV. Both the energy location of the bumps and
the strength contained agree with the features
observed in Fig. \ref{fig3}. Similarly, the B(GT$^+$) distributions in 
Ref. \cite{suhonen14} and in the corresponding Fig. \ref{fig4} in this
work, contain a similar amount of strength concentrated in the very
low-lying excitation energy.

Figs. \ref{fig5} and \ref{fig6} show the results for $^{150}$Nd 
and $^{150}$Sm, respectively.
We compare the QRPA results from SLy4 for prolate and oblate shapes with
the data from $(^3$He,$t)$ on $^{150}$Nd and from $(t,^3$He) on $^{150}$Sm 
in Ref. \cite{guess11}.
In general terms, we reproduce fairly well the global properties of the
GT strength distributions, including the location of the GT resonance
and the total strength measured, although the experimental GT resonances 
appear more fragmented than the calculated ones.
Similar comments to the previous case $A=116$ relative to the contributions
from GT and IVSM components apply now to the case $A=150$. Microscopic 
QRPA calculations with residual interactions obtained from the CD-Bonn 
potential were performed in Ref. \cite{guess11} to estimate the GT and
IVSM components of the strength B(GT+IVSM) measured. From this analysis 
it was also concluded that the strength B(GT$^-$+IVSM$^-$) below 20 MeV 
correspond to a large extent to GT strength, whereas the measured 
strength B(GT$^+$+IVSM$^+$) below 3 MeV could be assigned to the 
GT component.

In Fig. \ref{fig7} we show the results for the GT$^\pm$ strength 
distributions obtained with three different Skyrme forces to see the
sensitivity of these distributions to the nucleon-nucleon effective
interaction used in the calculations. We can see that the profiles
obtained with the three forces are quite similar. The GT$^-$ resonances 
appear at close energies and contain about the same strength for the
three interactions. The GT$^+$ strength is very small compared to the
GT$^-$ and is concentrated for the three forces in the excitation energy 
region below 5 MeV, with a bump in  $^{150}$Sm centered at about 15 MeV.

Figures \ref{fig8}, \ref{fig9}, \ref{fig10}, and \ref{fig11} show in more 
detail the GT strength distributions with SLy4 in the low excitation-energy 
range. In the case of $^{116}$Cd (Fig. \ref{fig8}) we have included, in 
addition, the experimental GT strength measured by ground-state to 
ground-state electron capture on $^{116}$In \cite{wrede13}. 
This value agrees well with the strength obtained with the prolate shape
and to a less extent with the oblate one, but it is too low compared to the 
spherical calculation. Similarly, in the case of $^{116}$Sn (Fig. \ref{fig9}) 
we show with an open circle the experimental GT strength measured by the
ground-state to ground-state $\beta^-$ decay ($^{116}$In 
$\rightarrow \, ^{116}$Sn). One should notice that
the B(GT) strength from Ref. \cite{rakers05} is calibrated to this value.
Although a detailed spectroscopy is beyond the capabilities of our model
and the isolated transitions are not well reproduced by our calculations,
the overall agreement with the total strength contained in this reduced
energy interval, as well as with the profiles of the accumulated strength
distributions is reasonable, especially for the prolate shapes in both 
$A=116$ partners in Figs. \ref{fig8} and \ref{fig9}, as well as with the 
prolate profiles in the $A=150$ partners displayed in  
Figs. \ref{fig10} and \ref{fig11}. 

\subsection{Double-beta decay}
\label{dbd}

It is well known that the $2\nu\beta\beta$ NMEs are very sensitive to the 
residual interactions, as well as to differences in deformation between 
initial and final nuclei \cite{alvarez04,simkovic04}.
To illustrate the dependence of the NMEs to the residual forces, we
show in Fig. \ref{fig12} the SLy4-QRPA results with prolate shapes 
for the $^{150}$Nd~$\rightarrow\,^{150}$Sm for various strengths of
the coupling constants  $\chi_{ph}^{GT}$ and $\kappa_{pp}^{GT}$.
The shaded region corresponds to the experimental NMEs extracted 
from the measured $2\nu\beta\beta$ half-lives, using bare
($g_A=1.273$) or quenched ($g_A=1$) values.
We can see that the experimental NMEs contained in the shaded
region are reproduced in windows of the parameters that include
the values used in this work ($\chi_{ph}^{GT}= 0.09$ MeV and 
$\kappa_{pp}^{GT}= 0.055$ MeV in $^{150}$Nd).

In the next figures we show the running sums for the $2\nu\beta\beta$ 
NMEs. These are the partial contributions to the NMEs of all the $1^+$
states in the intermediate nucleus up to a given energy.
We see in Fig. \ref{fig13} the running sums for the $2\nu\beta\beta$
decay in $^{116}$Cd for three combinations of the initial and final
nuclear shapes. The NMEs of crossing deformations produce
much lower values than the experimental window and are not shown.
Similarly, we show in Fig. \ref{fig14} the running sums for the 
$^{150}$Nd~$\rightarrow\,^{150}$Sm decay using prolate shapes for
both nuclei. 
Results obtained with the energy denominator $D_1$ (\ref{den1}) are 
displayed with thin lines, whereas results obtained with $D_2$ 
(\ref{den2}) are shown with thick lines. We can see that $D_2$ 
denominators produce larger NMEs than  $D_1$. The main difference
is originated at low excitation energies, where the relative
effect of using shifted energies is enhanced. The effect at larger 
energies is negligible and we get a constant difference between
$D_1$ and $D_2$, which is the difference accumulated in the first 
few MeVs. We also observe negative contributions coming from the
region of the GT resonances at about 15 MeV. This feature was studied 
in Ref. \cite{fang10} and was related to the strength of the $pp$ 
residual interaction.
The small contribution to the $2\nu\beta\beta$ NMEs from the region 
of the GT resonance is due to the joint effects of large energy 
denominators in Eq. (\ref{mgt}) and the mismatch between the branches 
of the GT$^-$ and GT$^+$ excitations in the energy region of the
GT resonance.

In Figs. \ref{fig15} and \ref{fig16} we have the $2\nu\beta\beta$ NMEs 
calculated with the denominator $D_2$ for three Skyrme forces in 
$^{116}$Cd and $^{150}$Nd, respectively. The results correspond to the 
prolate shapes in both initial and final nuclei. 
The sensitivity of the NMEs to the Skyrme interaction is manifest
in the figures, but it is not as significant as the sensitivity
found to other aspects such as the residual interactions, the deformation, 
or even the energy denominators.

Another issue worth to comment is the treatment of the quenching factor 
of the axial-vector coupling constant $g_A$.
The physical reasons for this quenching have been studied elsewhere 
\cite{osterfeld92,bertsch82,caurier05} and are related to the role of 
non-nucleonic degrees of freedom, absent in the usual theoretical models, 
and to the limitations of model space, many-nucleon configurations,
and deep correlations missing in these calculations.
The implications of this quenching on the description of single-$\beta$
and double-$\beta$ decay observables have been considered recently in
several works \cite{suhonen14,suhonen13,fogli08,barea13,caurier12}.
In those works both the effective value of $g_A$ and the coupling strength
of the residual interaction in the particle-particle channel are considered
free parameters of the calculation. The striking result found 
is that very strong quenching values are needed to reproduce simultaneously
the observations corresponding to the $2\nu\beta\beta$ half-lives and 
to the single-$\beta$ decay branches, namely
$\log ft (EC)$ values for the ground state to ground state decay of the
intermediate nucleus into the initial one, and the 
$\log ft (\beta^-)$ values for the ground state to ground state decay of the
intermediate nucleus into the final one. 
Different procedures for adjusting those parameters have been
explained in Refs. \cite{suhonen14,suhonen13,fogli08,barea13,caurier12}. 
The most recent analysis performed in Ref. \cite{suhonen14} within a QRPA 
formalism comes to the conclusion that considerable quenching is needed, 
with an average value 
$\langle g_A \rangle \approx 0.6 \pm 0.2\, (q=0.47 \pm 0.16)$.
The need of a strong quenching is also required in other theoretical
approaches, such as different QRPA calculations 
($q=0.6-0.7$) \cite{fogli08}, IBM-2 models \cite{barea13}, where the quenching 
obtained is $g_A=0.41 \, (q=0.32)$ for $^{116}$Cd and
$g_A=0.35 \, (q=0.28)$ for $^{150}$Nd, or shell-model calculations
that found quenching factors in the interval $q=0.45-0.74$ \cite{caurier12}
for double-$\beta$ emitters from $^{48}$Ca up to $^{150}$Nd.

One should note that the QRPA calculations leading to the strong quenching 
that fits the $2\nu\beta\beta$ NMEs have been performed within a 
spherical formalism neglecting possible effects from deformation degrees 
of freedom. Because the main effect of deformation is a reduction of the 
NMEs, deformed QRPA calculations shall demand less quenching to fit the 
experiment.
Although the goal of this paper is not to extract consequences on the 
quenching needed to fit the experiment, 
it will be very interesting to explore in the future the 
implications of deformation.
This study will be part of a more ambitious project aiming to extract
an effective $g_A$ value that fits the decay observables taking into account
the uncertainties related to the theoretical description and including
all possible $2\nu\beta\beta$-decay candidates.

\section{Summary and Conclusions}
\label{sec:conclusions}

In summary, using a theoretical approach based on a deformed HF+BCS+QRPA
calculations with effective Skyrme interactions, pairing correlations, 
and spin-isospin residual separable forces in the $ph$ and $pp$ channels,
we have studied simultaneously the GT strength distributions of the
double-$\beta$ decay partners ($^{116}$Cd, $^{116}$Sn) and ($^{150}$Nd, $^{150}$Sm),
reaching the intermediate nuclei $^{116}$In and $^{150}$Pm, respectively, as
well as their  $2\nu\beta\beta$ NMEs.
Our results for the energy distributions of the GT strength
have been compared to recent data from charge-exchange reactions,
whereas the calculated $2\nu\beta\beta$ NMEs have been compared with 
the experimental values extracted from the measured half-lives for 
these processes.

We have discussed the sensitivity of our results to the various ingredients
in the theoretical formalism. Namely, to the Skyrme force, to the residual 
interactions, to deformation, to effective $g_A$ values, and to the 
treatment of the energy denominators of the NMEs.
We found different sensitivities to them that have been analyzed and discussed.
All in all, the method used in this work has demonstrated to be well suited to 
account for the rich variety of experimental information available on the nuclear 
GT response. The global properties of the energy distributions of the GT strength
and the $2\nu\beta\beta$ NMEs are reasonably well reproduced, the exception is
the detailed description of the low-lying GT strength distributions that could
clearly be improved.
It will be interesting in the future to extend these calculations by including
all the double-$\beta$ decay candidates and to explore systematically the
potential of this method.

\section*{Acknowledgments} 

We are grateful to E. Moya de Guerra for useful discussions.
This work was supported in part by MINECO (Spain) under Research Grant
No.~FIS2011-23565 and by Consolider-Ingenio 2010 Programs CPAN 
CSD2007-00042.


\end{document}